# Public perceptions of AI-driven decision-making in healthcare: A structural equation modeling approach

Leonie Westerbeek[1], Ernesto de León[1], Julia CM van Weert[1]

[1]Amsterdam School of Communication Research/ASCoR, University of Amsterdam, Amsterdam, The Netherlands

Correspondence to Leonie Westerbeek, Amsterdam School of Communication Research, University of Amsterdam, Nieuwe Achtergracht 166, 1018 WV Amsterdam, The Netherlands; l.westerbeek@uva.nl; Tel: +31 (0)20 525 3680

**Objectives:** Artificial intelligence (AI) is increasingly integrated into healthcare to support diagnostics, decision-making, and administrative processes, offering potential gains in efficiency, accuracy, and health communication. However, the successful implementation of AI depends not only on technical performance but also on public perceptions of its helpfulness, riskiness, and fairness. This study examines public perceptions of automated decision-making (ADM) in healthcare.

**Methods:** Data were drawn from the first wave of an ongoing six-wave longitudinal survey panel. The final sample consisted of 3,915 respondents (mean age 52.1 years; 52.1% female), and was analyzed with structural equation modelling (SEM). Perceptions of ADM in healthcare as *helpful*, *risky*, and *fair* were treated as the key dependent variables. AI literacy, familiarity with different forms of AI, confidence in clinicians' ability to distinguish AI- from human-generated content, use of conversational agents for health information, and use of traditional digital health information sources were included as exogenous. Model fit was assessed and improved based on modification indices.

**Results:** Greater familiarity with different forms of AI, higher confidence in the clinician's ability to recognize AI-generated content, and use of conversational agents for health information were associated with greater perceived *helpfulness*. Use of conversational agents was associated with lower perceived *risk*, whereas greater familiarity with AI and greater reliance on traditional health information sources were associated with higher perceived *risk*. Perceptions of ADM as *fair* were most strongly predicted by confidence in the clinician's ability, with additional small positive associations with AI familiarity, AI literacy, and use of conversational agents.

**Conclusions/practice implications:** Public perceptions of ADM in healthcare are shaped by technological familiarity, use of conversational agents, and confidence in human oversight.

Overall, ADM's perceived helpfulness and fairness are driven more by trust in healthcare professionals than by trust in the technology itself.

## 1. Introduction

Artificial Intelligence (AI) is playing an increasingly important role in healthcare, enhancing healthcare processes such as diagnostics, clinical decision-making, and administrative tasks [1]. These applications have the potential to make healthcare not only more efficient but also more accessible and equitable [2]. AI-driven technologies have the potential to improve efficiency, reduce human error, and enhance patient outcomes [3]. Beyond these operational improvements and enhancements of overall healthcare quality, AI also has the potential to enhance health communication by facilitating more effective interactions between patients and clinicians [4,5]. AI-driven systems are inevitably becoming a part of patient-provider interactions, influencing how health information is accessed, interpreted, and acted upon [6,7]. As such, the integration of AI into healthcare is not solely a technological innovation, but also a communicative shift that comes with opportunities and challenges for health communication.

However, the promise of AI in healthcare is accompanied by concerns from the public. Algorithms do not inherently understand context, emotion, or other human nuances [8]. Consequently, there is a risk that over-reliance on AI systems could depersonalize care and reduce opportunities for empathetic interaction [9]. Additionally, matters related to algorithmic bias, data privacy, and transparency have raised ethical concerns about the fairness and inclusivity of AI in healthcare and health communication [10]. These concerns highlight that the success of AI in healthcare cannot be evaluated solely on technical performance, but must also consider how these systems are perceived by those affected by them.

In this context, public perceptions play a critical role. The effectiveness of AI in fostering better communication and healthcare in general depends not only on its technical effectiveness, but also largely on public perceptions of its helpfulness, riskiness, and fairness [11]. A technology that is seen as helpful, trustworthy, and fair is more likely to be embraced and integrated into daily healthcare practices [12,13]. Conversely, skepticism or fear surrounding AI can lead to resistance, underutilization, or outright rejection, regardless of the technology's objective performance [14]. Yet, despite the growing prevalence of AI-driven systems in healthcare, public perceptions toward these technologies remain underexplored.

Understanding these public perceptions is especially important as both patients and healthcare providers increasingly often interact with AI-driven health tools. Patients interact with tools such as online symptom checkers, digital health platforms, or conversational agents embedded in healthcare systems [15,16]. However, the extent to which individuals actually use these digital health information sources varies considerably, which may in turn shape how they think about AI in healthcare more broadly. Healthcare providers, for instance, use AI in the process of clinical decision-making, resulting in an AI-driven decision-making process [17]. While these automated decision-making (ADM) systems can complement human expertise by providing data-driven insights and reducing cognitive burden, they also introduce new complexities. The interplay between algorithmic recommendations and human judgement may alter traditional clinical decision-making structures [18]. The presence of ADM systems raises questions about how much trust healthcare providers, patients, and the general public should place in algorithmic outputs. As such, understanding how the public perceives the role of ADM systems in healthcare is crucial. Not only for the successful implementation of these technologies, but also for safeguarding patient-centered care in a rapidly evolving digital landscape.

Understanding how the public perceives and evaluates ADM systems in healthcare is essential for healthcare providers and policymakers to develop communication strategies that emphasize AI's supportive role, address concerns about risks and fairness, and ensure that AI integration enhances rather than disrupts the patient-clinician relationship. Public perceptions of ADM systems in healthcare are versatile; they are shaped by a range of factors, such as prior exposure to and familiarity with AI, and AI literacy [19]. AI literacy entails the extent to which individuals are capable of critically evaluating and effectively using AI technologies [20]. In addition, individuals' use of digital health information sources, and more specifically their use of conversational agents (e.g., chatbots) as a source of health information, may shape how they perceive ADM systems, as such experiences provide direct exposure to AI-mediated health communication. Importantly, individuals' confidence in clinicians' ability to differentiate between AI- and human-generated content also plays a role in shaping public perceptions. If patients believe that healthcare professionals cannot adequately evaluate or override AI recommendations, this could negatively impact their perception of the technology, but also of the broader healthcare system [21].

This study aims to assess how the general public perceives the helpfulness, riskiness, and fairness of ADM systems in healthcare. Using a structural equation modelling approach, we explore the relationships between these perceptions and several contextual factors. Ultimately, this research contributes to a growing body of literature at the intersection of technology and health communication. By understanding public perceptions of ADM systems in healthcare, we can ensure that their integration into the healthcare system enhances, rather than undermines, the communication processes and overall healthcare quality.

## 2. Methods

### 2.1 Design and Sample

Data collection took place through the AlgoSoc Longitudinal Survey Panel: a three-year data collection effort aimed at understanding public perceptions of algorithms and their relationship to the areas of politics, media, health, and justice [22]. In total, six waves of data will be collected through this panel. This paper describes the analysis of data from the first wave. The survey was fielded within Centerdata's LISS Panel, a probability-based panel of 5000 Dutch households, to obtain a representative sample of the Dutch population. Ethical approval of the study was granted by the institution's ethical review board (ref: FMG-7010). The dataset originally contained 4017 respondents. Because of missing values, 102 respondents were lost. This resulted in a final sample of 3915 respondents.

**2.2 Measures**

The endogenous variables, respondents' perception of ADM in healthcare being *helpful*, *risky*, and *fairer*, were each measured with 1 item on a 7-point Likert scale (1 = strongly disagree, 7 = strongly agree). The items were phrased as *more automated decision-making in healthcare will be helpful*, *more automated decision-making in healthcare will be risky*, and *more automated decision-making in healthcare will be fairer*.

AI literacy was measured with 5 items on a 7-point Likert scale (1 = strongly disagree, 7 = strongly agree) based on the DigIQ validated scale [23]. Items included, for instance, *I recognize it when a website or app uses AI to tailor the content to me*, and *I know how to access the data that AI systems use to tailor content to me*. These five items were grouped together in a mean scale ($\alpha = .90$) that represents respondents' level of AI literacy.

AI familiarity was also measured with 5 items on a 7-point Likert scale (1 = not at all familiar, 7 = very familiar). Respondents were asked: *How familiar are you with the following?* They were then provided with the following five items: *Generative AI*, *ChatGPT*, *Deepfakes*, *Algorithms*, and *Chatbots*. These five items were grouped together in a mean scale ($\alpha = .91$) that represents respondents' familiarity with different forms of AI.

Respondents' usage of more traditional digital health information sources was measured with 4 items on a 7-point scale. Respondents were asked: *How often have you used the following resources for health purposes?* They were then provided with the following four items: *Search engines; Official, general health website; Patient portal; Official website* *of a healthcare institution.* These four items were grouped together in a mean scale (α = .77) that represents respondents' use of more traditional digital health information sources.

Respondents' usage of conversational agents as a source of health information was measured on a 7-point scale (1 = never, 7 = always in case of a health question) with 1 item: *How often have you used the following resources for health purposes? Chatbot or virtual assistants (such as ChatGPT, Bard/Gemini, Bing Chat)*.

Lastly, respondents' perception of clinicians' ability to differentiate between AI-generated versus human-generated content was measured on a 7-point Likert scale (1 = not sure at all, 7 = very sure) by asking: *How confident are you that the following groups can tell the difference between content created by AI and content created by humans? Physicians*.

**2.3 Model Specification**

Structural equation modelling was performed using AMOS 29. All variables in the model were included as observed variables. The model included respondents' perception of ADM in healthcare as being *helpful*, *risky*, and *fairer* as endogenous variables. AI literacy, familiarity with different forms of AI, perceptions of clinicians' ability to differentiate between AI versus human-generated content, respondents' usage of conversational agents as a source of health information, and respondents' usage of more traditional digital health information sources were included as exogenous variables. Respondents' age, sex, and level of education were included in the model as control variables. The model is depicted in Figure 1. Multivariate non-normality was detected for respondents' usage of conversational agents as a source of health information (kurtosis > 5). A log transformation was therefore applied to this variable. Kurtosis remained high after transforming; therefore, p-values of direct effects were based on bootstrapped 95 % bias-corrected confidence intervals [24]. Exogenous variables were allowed to correlate freely. The model fit was evaluated using three indicators: Chi-Square (*p*-value should be >.05), RMSEA (ideally <.05; 0.05-0.08 indicates acceptable fit), and CFI (ideally >.95; >.90 indicates acceptable fit). The model fit was evaluated and improved based on the modification indices. The final model had an acceptable fit. The RMSEA (.074) and CFI (.911) scores were acceptable. The chi-square was significant, which is undesirable. However, the chi-square test is highly sensitive when used in SEM with large samples, often yielding significant results even for trivial discrepancies [25]. Given our large sample size (*n* = 3915) and the acceptability of the other indices, we decided to continue with this model.

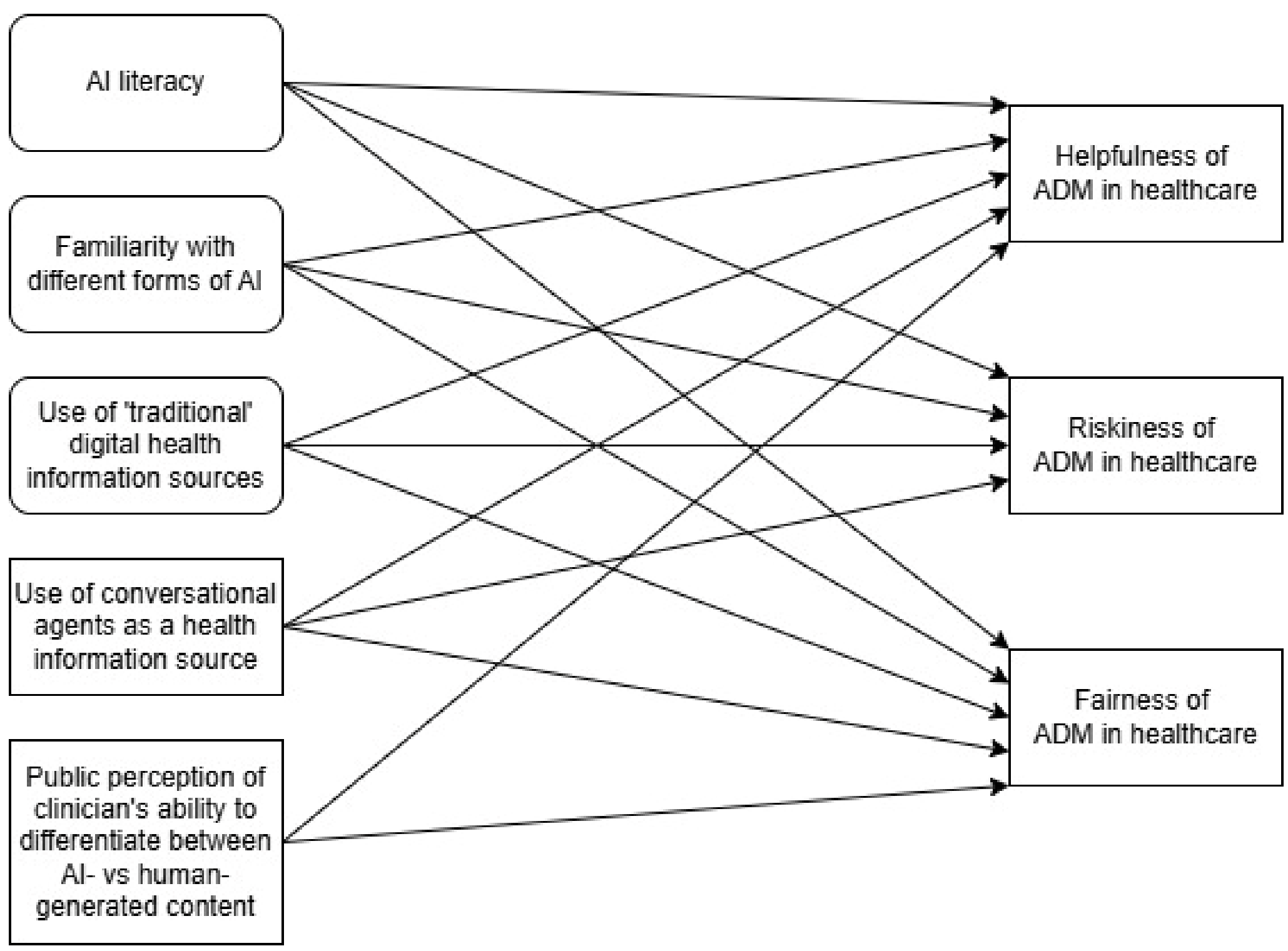

**Figure 1.** Tested model (simplified, without covariances and error correlations).

## 3. Results

The final sample consisted of 3915 respondents, with a mean age of 52.1 years (*SD* = 18.6). The sample was nearly balanced in terms of gender, including 2040 female (52.1%) and 1875 male respondents. On average, respondents reported relatively low levels of AI literacy (*M* = 2.60, *SD* = 1.43) and limited familiarity with different forms of AI, such as generative AI, ChatGPT, and deepfakes (*M* = 2.97, *SD* = 1.72). Respondents also reported modest use of traditional digital health information sources, such as search engines, official health websites, and patient portals (*M* = 2.73, *SD* = 1.17), and they rarely used conversational agents like ChatGPT or Gemini as a source of health information (*M* = 1.33, *SD* = 0.96). In addition, respondents were, on average, not very confident in physicians' ability to distinguish AI-generated from human-generated content (*M* = 3.13, *SD* = 1.55).

Regarding their perceptions of ADM in healthcare, respondents were, on average, most inclined to view ADM as risky (*M* = 4.82, *SD* = 1.66). Perceptions of ADM as helpful were more neutral (*M* = 3.51, *SD* = 1.60), while respondents were least likely to perceive ADM as fair (*M* = 3.10, *SD* = 1.50). Together, these descriptive results suggest that the public tends to associate ADM in healthcare more strongly with risk than with helpfulness or fairness.

**3.1 Helpful**

As depicted in Table 1 and Figure 2, four exogenous variables were significantly associated with perceiving more ADM in healthcare as helpful. First, the model showed a significant, positive relationship between familiarity with different forms of AI and helpfulness. This means that respondents who were more familiar with different forms of AI perceived ADM as more helpful than participants with lower familiarity scores. The model also showed a significant, positive relationship between perceptions of clinicians being able to differentiate between AI-generated content vs. human-generated content and helpfulness. If participants were more confident that clinicians can differentiate between AI- versus human-generated content, they were more likely to perceive ADM in healthcare as helpful. We also found a small, significant, positive relationship between AI-literacy and helpfulness. If respondents had higher AI-literacy scores, they were also more likely to perceive ADM in healthcare as helpful. Lastly, the model showed a small, significant, positive relationship between the use of conversational agents and helpfulness. The more respondents use conversational agents as a source for health information, the more they perceive ADM in healthcare to be helpful.

**3.2 Risky**

As shown in Table 1 and Figure 2, three exogenous variables were significantly associated with perceiving more ADM in healthcare as risky. Using conversational agents as a source for health information had a significant, negative relationship with riskiness. Respondents who indicated they used chatbots or virtual assistants as a health information source perceived ADM in healthcare as less risky. Furthermore, the model shows a small, significant, positive relationship between familiarity with different forms of AI and riskiness. Respondents who scored high on familiarity with AI perceived ADM in healthcare as more risky. Lastly, we found a small, significant, positive relationship between the use of traditional health information sources and riskiness. The more respondents used traditional health information sources, the more likely they were to perceive ADM in healthcare as risky.

**3.3 Fairer**

As depicted in Table 1 and Figure 2, four exogenous variables were significantly associated with perceiving more ADM in healthcare as fairer. The strongest, significant, positive relationship was found between confidence that clinicians can differentiate between AI-generated versus human-generated content and fairness. If respondents were more confident that clinicians can differentiate between AI- versus human-generated content, they were more likely to perceive ADM in healthcare as fairer. The model also showed a small, significant, positive relationship between familiarity with different forms of AI and fairness. Respondents who scored high on familiarity with AI perceived ADM in healthcare as fairer. The same goes for the relationship between using conversational agents as a source for health information and fairness. Respondents who indicated they used conversational agents as a health information source perceived more ADM in healthcare as fairer. Lastly, a small, significant, positive relationship was found between AI-literacy and perceived fairness. Respondents with a higher AI-literacy perceived more ADM in healthcare as fairer.

**Table 1**
*Standardized effects*

| | More ADM is ***helpful*** | More ADM is ***risky*** | More ADM is ***fairer*** |
|---|---|---|---|
| **AI-literacy** | .047* | -.017 | .048* |
| **Familiarity with AI** | .184*** | .068** | .081** |
| **Clinicians' perceived ability to differentiate between AI- vs. human-generated content** | .122*** | | .115*** |
| **Traditional digital health information sources** | .027 | .038* | .006 |
| **Conversational agents as a source for health information** | .039* | -.106*** | .083*** |

*Note.* * $p < .05$, ** $p < .01$, *** $p < .001$

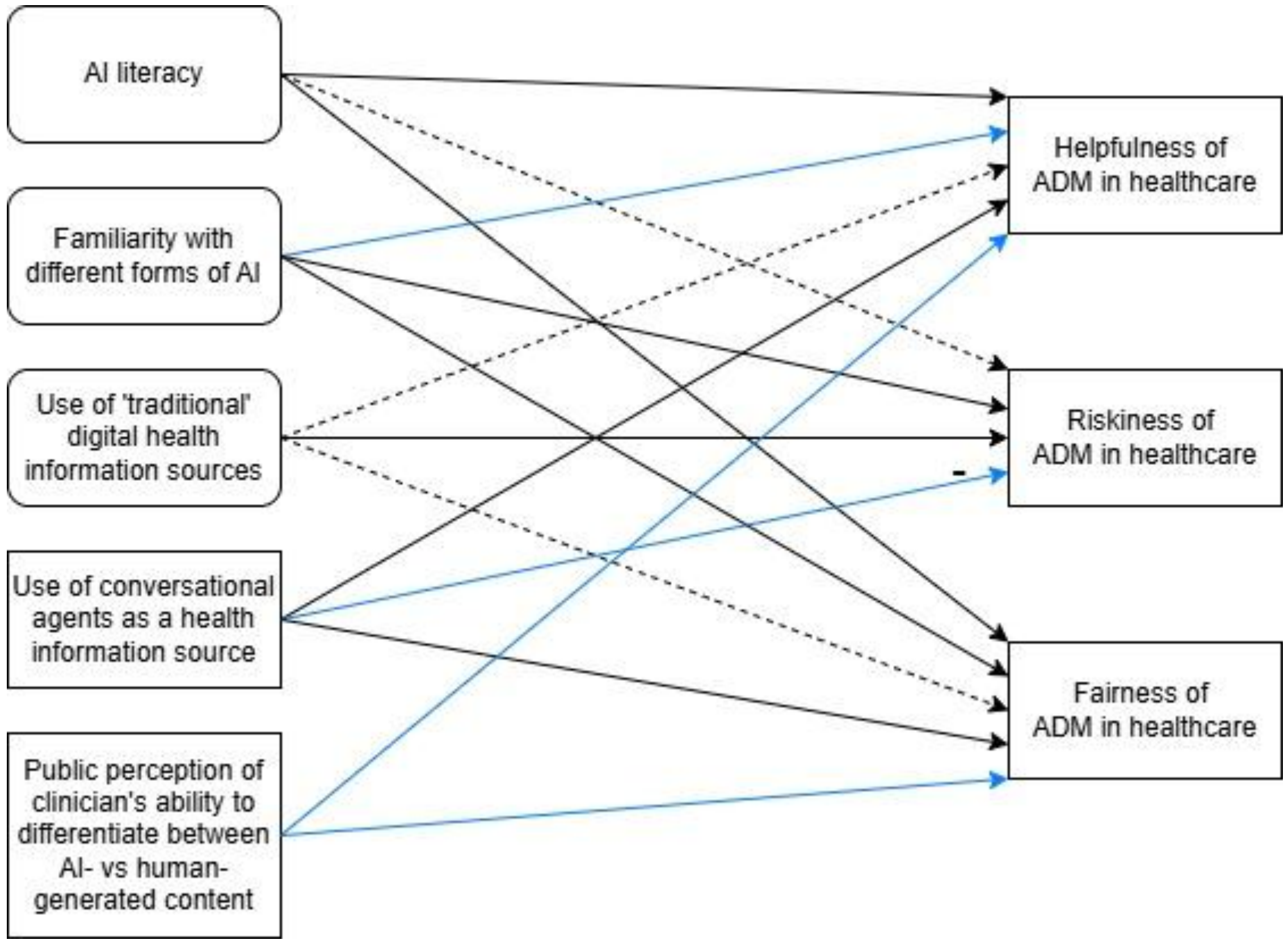

**Figure 2.** Model results

*Note.* Solid blue arrows indicate a statistically significant ($p < 0.05$) relationship with a standardized regression weight >.1. Solid black arrows indicate a significant relationship with a standardized regression weight <.1. Dashed arrows indicate non-significant relationships.

## 4. Discussion and Conclusion

### 4.1 Findings and Implications

This study examined how the general public perceives ADM systems in healthcare, focusing on perceptions of its helpfulness, riskiness, and fairness. We assessed how these perceptions are shaped by AI literacy, familiarity with AI, use of conversational agents for health information, use of traditional health information sources, and confidence in clinicians' ability to differentiate between AI- and human-generated content. The findings reveal a pattern of associations, showing that public perceptions of ADM are related to both technological familiarity and usage, and confidence in human oversight.

Across all three outcome variables, familiarity with different forms of AI emerged as a statistically significant predictor. Respondents who were more familiar with AI technologies were more likely to perceive ADM in healthcare as helpful and fair, but also as risky. This suggests that familiarity does not simply translate into uncritical acceptance. Rather, greater familiarity with different forms of AI may increase awareness of both its potential benefits and its limitations. This supports previous literature, suggesting that familiarity relates to overall acceptance of AI applications [26], but in addition suggests that it increases perceived risks. This finding implies that increasing familiarity with AI should not be expected to eliminate concerns about its risks, but could facilitate more realistic and informed evaluations.

AI literacy showed a smaller but still positive association with perceptions of helpfulness and fairness, while it was not significantly related to perceived riskiness. This suggests that understanding how AI systems function and how they use data may primarily help recognize their potential helpfulness and fairness, but is not related to their perceived riskiness. It is surprising that AI literacy does not seem to be related to perceived riskiness in our model. Previous literature does suggest that a higher AI literacy could mitigate perceived risks, such as privacy concerns [27]. Another study found positive associations between AI literacy and positive attitudes, but no correlations between AI literacy and negative attitudes, in line with our findings [28]. Perhaps the concept of AI literacy is difficult to grasp with self-reported measures, or the measure we used was not elaborate enough. It could also be that higher AI literacy might make some people very aware of its risks, while causing others to become overconfident, thereby mitigating risks. This might make it difficult to find an association between AI literacy and its perceived riskiness. Overall, initiatives aimed at improving AI literacy may be effective in strengthening perceptions of helpfulness and fairness, but should be complemented with transparent communication about, for instance, bias, transparency, and privacy, to address risk perceptions more directly.

One of the most notable findings concerns the role of clinicians' perceived ability to differentiate between AI- and human-generated content. This perceived ability was positively associated with perceptions of helpfulness and fairness. These results highlight the central role of the clinician in shaping public perceptions of ADM in healthcare [29]. Perhaps, when clinicians are seen as capable of identifying AI-generated outputs, they are also perceived as being able to effectively oversee and critically evaluate these systems. In this way, perceived ability to detect AI functions as a proxy for meaningful human oversight, highlighting its central role in shaping public perceptions of ADM in healthcare. This is in line with previous literature, also suggesting that human oversight plays a central role in recognizing the benefits of AI usage in healthcare [30]. Our findings suggest that the AI literacy of the healthcare professional using an ADM system is more critical than the AI literacy of the patients themselves. This reinforces the idea that public perceptions of ADM in healthcare are fundamentally relational: In the end, patients value the clinician's perceived ability to work with these systems more than their own knowledge of the technology.

The use of conversational agents as a source of health information also played a meaningful role. Respondents who used chatbots or virtual assistants as a health information source more frequently perceived ADM as more helpful and fair, and less risky. This suggests that personal experience with AI-driven health tools may normalize algorithmic involvement in healthcare and reduce perceived riskiness, in line with previous research [31]. Regular users of conversational agents may develop practical expectations of what such systems can and cannot do, potentially leading to lower perceived risk. At the same time, the relatively low mean usage of conversational agents in the sample indicates that such experiences remain limited for most respondents, mirroring broader findings that conversational agents are not yet routinely used in healthcare contexts [32]. As these tools become more widespread, their influence on public perceptions of ADM in healthcare is likely to grow.

Lastly, the use of more traditional digital health information sources (e.g. patient portals or official healthcare institution websites) only showed a very small association with perceived riskiness. The use of more conventional digital sources for health information was related to perceiving ADM in healthcare as riskier. Use of traditional digital health information sources was not related to perceived helpfulness and fairness, implying that this type of health information-seeking behavior is less relevant for predicting public perceptions of ADM in healthcare.

**4.2 Strengths and Limitations**

A key strength of this study is its large, nationally representative sample drawn from the LISS Panel. This enhances the generalizability of the findings to the Dutch population and provides a robust empirical basis for examining public perceptions of ADM in healthcare. Additionally, the use of structural equation modeling allowed for the simultaneous examination of multiple predictors and outcome variables, providing a comprehensive view of how different factors relate to perceptions of helpfulness, riskiness, and fairness.

Despite these strengths, a few limitations should also be acknowledged. First, the cross-sectional nature of the data precludes causal inference. While the model identifies significant associations, it cannot determine their directionality. For example, while greater familiarity with AI is associated with higher perceived helpfulness, it is also possible that individuals who already view ADM positively are more inclined to engage with or learn about AI. Longitudinal data from future waves of the AlgoSoc panel could help disentangle these dynamics.

Second, the key outcome variables (i.e. perceived helpfulness, riskiness, and fairness) were each measured with a single item. Although single-item measures are not uncommon in survey research and can be appropriate for broad evaluative judgments, they limit measurement precision and reliability. Future studies could benefit from multi-item scales that capture different dimensions of these public perceptions.

Third, while this study included several relevant predictors, it did not account for other factors that may shape perceptions of ADM, such as personal health status, prior experiences with the healthcare system, or broader trust in institutions. Additionally, cultural and contextual factors specific to the Dutch healthcare system may influence the findings, limiting their applicability to other national contexts with different healthcare infrastructures or regulatory frameworks.

**4.3 Future Research**

Future research could build on the present findings in several ways. Longitudinal analyses using subsequent waves of the AlgoSoc panel will be particularly valuable for examining how public perceptions of ADM evolve over time, especially as AI-driven systems become more integrated into healthcare practice. AI is a rapidly evolving field, and the other variables (e.g. AI literacy, use of conversational agents etc.) are also likely to evolve in subsequent waves. Longitudinal analyses will provide an insight into how these factors evolve and how this affects the relationships between these variables and the public perception of ADM in healthcare .

Another interesting avenue for future research could involve comparative perspectives. It would be interesting to explore how differences in healthcare systems, regulatory environments, and cultural attitudes toward technology shape public perceptions of ADM. Additionally, comparing public perceptions with those of specific patient groups and healthcare professionals could reveal alignment or discrepancies among stakeholder groups.

Finally, qualitative research could complement our survey-based approach by exploring public perceptions of ADM in healthcare more in-depth. Interviews or focus groups could provide richer insights into how people interpret algorithmic decision-making, how they envision its role in clinical encounters, and what conditions they consider necessary for its safe use.

**4.4 Conclusion and Practice Implications**

This study shows that public perceptions of ADM in healthcare are shaped mainly by AI familiarity, use of conversational agents for health information, and trust in human oversight. ADM is more likely to be perceived as helpful and fair when individuals are familiar with AI technologies, possess higher AI literacy, and, most crucially, believe that clinicians can critically engage with and control algorithmic outputs. At the same time, greater familiarity with AI may also heighten awareness of potential risks. These findings suggest that the successful integration of ADM into healthcare depends not only on technological performance, but also on effective communication about the role of AI and the continued centrality of clinicians in the decision-making process. Ultimately, public perceptions of the technology's helpfulness and fairness appear to rely more on confidence in the healthcare professional who uses it, rather than on trust in the technology itself.

**Funding Statement:**

This material is produced as part of AlgoSoc, a collaborative 10-year research program on public values in the algorithmic society which is funded by the Dutch Ministry of Education, Culture and Science (OCW) as part of its Gravitation programme (project number 024.005.017). Any opinions, findings, and conclusions or recommendations expressed in this material are those of the author(s) and do not necessarily reflect the views of OCW or those of the AlgoSoc consortium as a whole.